\documentclass{esannV2}

\usepackage{amssymb}
\usepackage{amsmath}
\usepackage{multirow}
\usepackage{subcaption}
\usepackage{hyperref}
\usepackage{cleveref}
\usepackage{pgfplots}
\usepackage{booktabs}

\begin{document}




\title{Dual Stream Graph Transformer Fusion Networks for Enhanced Brain Decoding}


\author{Lucas Goené and Siamak Mehrkanoon
%
%
\vspace{.3cm}\\
%
Department of Information and Computing Sciences, Utrecht University,\\
Utrecht, The Netherlands
}




\maketitle

\begin{abstract}

This paper presents the novel Dual Stream Graph-Transformer Fusion (DS-GTF) architecture designed specifically for classifying task-based Magnetoencephalography (MEG) data. In the spatial stream, inputs are initially represented as graphs, which are then passed through graph attention networks (GAT) to extract spatial patterns. Two methods, TopK and Thresholded Adjacency are introduced for initializing the adjacency matrix used in the GAT. In the temporal stream, the Transformer Encoder receives concatenated windowed input MEG data and learns new temporal representations. The learned temporal and spatial representations from both streams are fused before reaching the output layer. Experimental results demonstrate an enhancement in classification performance and a reduction in standard deviation across multiple test subjects compared to other examined models.

\end{abstract}







\section{Introduction}
\label{sec:introduction}
Electroencephalogram and Magnetoencephalography (MEG) are non-invasive methods used to explore neuronal activity, providing insights into the functioning human brain. The integration of deep machine learning techniques has led to the development of advanced data-driven models aimed at uncovering the underlying patterns of EEG and/or MEG data. Despite extensive research on human activity classification using EEG data \cite{paper:EEGNet, zhang2018cascade}, studies utilizing MEG data remain relatively scarce. MEG signals, while more costly and complex to acquire compared to EEG, offer greater precision owing to their superior spatio-temporal resolution. Recent studies have made significant progress in developing EEG decoding networks, that can analyze EEG data to decode information about brain activity. While EEG decoding has seen notable advancements, the application of similar techniques to MEG signals has been relatively less explored and developed. In recent literature, several deep convolutional neural network (CNN)-based models have been developed for analyzing both EEG \cite{paper:EEGNet, paper:kazatzidis2023novel} and MEG data \cite{abdellaoui2021enhancing}. Additionally, graph convolutional networks and Attention based networks have proven successful in decoding EEG signals \cite{paper:EEG-GAT, paper:DPGAT}, although they are less commonly applied to MEG data \cite{paper:MEG-GNN}. In some instances, existing architectures designed for EEG data are modified to accommodate MEG data, as evidenced by studies such as \cite{paper:MEG-EEGNet}. The authors in \cite{abdellaoui2021enhancing}, introduced AA-EEGNet as well as AA-CascadeNet by incorporating attention mechanisms, self and global attention, into both core EEGNet \cite{paper:EEGNet} and CascadeNet \cite{zhang2018cascade} models. Here, we develop a novel dual stream graph transformer fusion architecture for decoding Magnetoencephalography (MEG) data. Thanks to the two-stream architecture, the model learns both spatial and temporal representations which are then concatenated prior to reaching the output layer. Each stream autonomously receives and processes its respective input representation, while the entire architecture is trained in an end-to-end manner.

\section{Preliminaries}
\label{sec:preliminaries}

\subsection{Transformer Encoder}
The temporal stream of our model employs a Transformer Encoder. Transformer based models have demonstrated effectiveness in capturing sequential dependencies in various tasks. In this process, the recording from each channel is initially divided into smaller overlapping segments, forming matrices of dimensions $(c \times d)$, where $c$ represents the number of channels and $d$ is the segment length. Each row of a matrix, i.e. $(c \times d)$, is subsequently inputted into the Transformer encoder, which generates a new embedding of the same dimension. Similar to \cite{paper:transformer}, here we also employ the multi-head self-attention mechanism. The self-attention mechanism calculates attention scores across the input sequence, enabling the model to determine the relevance of each element to others. By utilizing multiple attention heads, the model captures various aspects of inter-element relationships, facilitating the capture of long-range dependencies. After computing attention scores, the Transformer generates new embeddings for each element, which then undergo processing through a feed-forward block. This block comprises two dense layers with ReLU activation and linear transformation, respectively. The final output of the Transformer encoder is produced by the last dense layer, encapsulating the learned representations of the input sequence.

\subsection{Graph Attention Module}
The spatial stream of our model incorporates a Graph Attention Module, utilizing the brain's structural relations. Here, each MEG channel is modeled as a node in a graph. The use of Graph Attention layers enables the extraction of spatial features. Following the lines of \cite{paper:GAT}, we also incorporate the multi-head approach, which results in specific attention values and weight matrices for each head. The outputs of the heads are aggregated using concatenation. These final embeddings form the output of the Graph Attention Module. 

\subsection{Adjacency Matrix - RBF Kernel}
\label{sec:rbf}
First the segments ($c \times d$) are divided in equally sized windows of size $w$, resulting in matrices of size $(c \times w)$. A graph is built for each input window with each MEG channel serving as a node. These nodes possess feature vectors composed of a sequence of consecutive time steps. To initialize the adjacency matrix of the graph for the spatial stream, we utilize a Radial Basis Function (RBF) Kernel. This kernel computes edge weights based on the physical distance between nodes (channels), allowing for a realistic representation of brain organization. The RBF kernel is defined as: $\textrm{RBF}(c_i, c_j) = \exp(-\gamma \cdot \|c_i - c_j\|^2)$, where $c_i$ and $c_j$ denote the location coordinates of their respective nodes, and $\gamma$ is a kernel bandwidth which controls the level of local flexibility in the model. Here, the kernel bandwidth, $\gamma = 100$, is determined empirically. We investigate three methods for initializing adjacency matrix: Fully Connected Adjacency (FC-Adj), Thresholded Adjacency (Thresh-Adj), and Top-K Adjacency (TopK-Adj). Each method produces a binary adjacency matrix. In fully connected adjacency, every possible edge is present in the matrix. In the other methods, a subset of edges is chosen based on their RBF value. With Thresh-Adj, only edges with a weight above a specified threshold are included. With TopK-Adj, for each node, only the top $K$ edges are selected to be included in the adjacency matrix.

\section{Proposed Model}
\begin{figure}[h]
    \centering
    \includegraphics[width=1\textwidth]{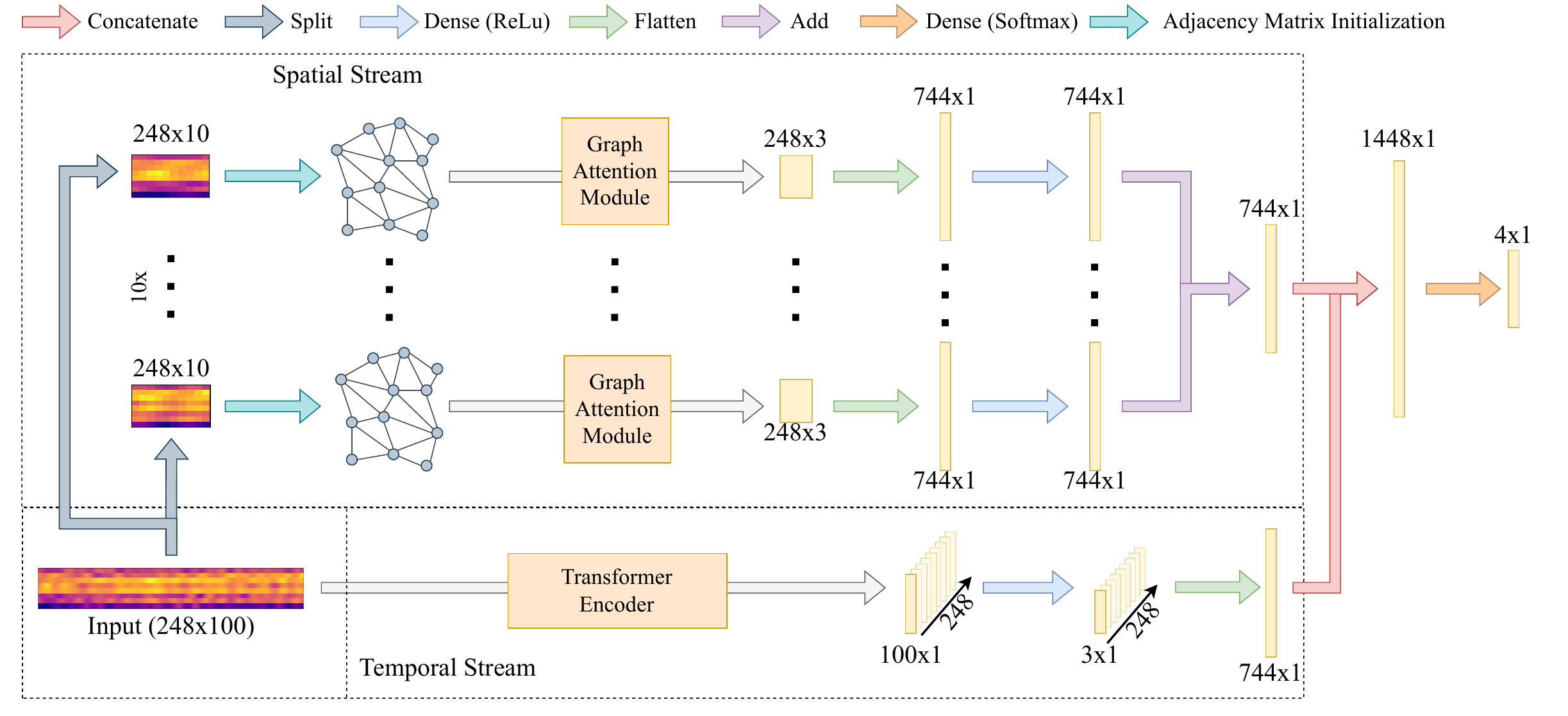}
    \caption{A visualization of our proposed DS-GTF architecture.  }
    \label{fig:model}
\end{figure}

In this paper we propose the Dual Stream Graph Transformer Fusion (DS-GTF) architecture, an adaptation of the AA-MultiviewNet proposed in \cite{abdellaoui2021enhancing}. The dual stream setup is used to extract both temporal and spatial features from the input data. Both streams receive a unique data representation as input, each derived from the same initial input source. The spatial stream uses a Graph Attention Module, where the inputs are presented as a graph, see Fig. \ref{fig:model}. The graphs are generated using one of the three discussed adjacency initialization methods. Next we pass each graph to a separate Graph Attention Layer, where each layer has three attention heads. The output of each GAT is flattened and passed through a dense layer. The output of all the dense layers are then added together to create the final embedding for the spatial stream. In the temporal stream each row of the overlapping segments of size $(c \times d)$ undergoes processing individually through the Transformer encoder. Here, the Transformer-based Encoder employs $8$ attention heads. Afterward, the output undergoes down-sampling through a dense layer and is flattened to match the output embedding of the spatial stream. The resulting outputs from both streams are then merged via concatenation, followed by a dense layer and a softmax layer. The implementation of the model is available on Github \footnote{\url{https://github.com/Lucasgoene/DS-GTF}}.

\section{Data Description}
\label{sec:data}
We utilize the MEG data sourced from the Human Connectome Project's 1200 Subjects Release (S1200) dataset, consistent with the dataset employed in \cite{abdellaoui2021enhancing}. The MEG recordings were conducted using a MAGNES 3600 system, comprising 248 magnetometer channels and 23 reference channels, with a sampling rate of $2034.5101$ Hz. The total number of subjects is $95$, but following the lines of \cite{abdellaoui2021enhancing} we only use $18$ subjects with reliable data. From this subset, $12$ subjects are allocated for training and validation purposes, while the remaining $6$ are designated for testing, following the same partitioning as in \cite{abdellaoui2021enhancing}. The tasks performed by the subjects during scanning encompassed four categories: \textbf{resting}, \textbf{story and math}, \textbf{working memory}, and \textbf{motor} tasks.

\section{Experiments}

Initially, we segment the recording from each channel into overlapping segments of length $d=100$ with a $50\%$ overlap. After segmentation, each segment undergoes normalization and is then split into 10 equally sized windows, each with a length of $w=10$. The values of $d$, overlap, and $w$ are tuned, with the reported values found empirically to be optimal. Consistent with \cite{abdellaoui2021enhancing}, we adopt a setup comprising twelve training subjects and six testing subjects. Model evaluation entails computing the average testing accuracy across the subjects in the testing set. We explore and optimize three distinct methods for initializing the adjacency graph, adjusting their respective parameters and comparing the best-performing model for each method. Optimization of the categorical cross-entropy loss is performed using the Adam optimizer, with a learning rate of $1e^{-4}$ and a batch size of $32$, both empirically found to be optimal. Every model was trained for $15$ epochs. 

\section{Results and Discussion}

The proposed DS-GTF model is assessed against the current state-of-the-art models for this dataset outlined in \cite{abdellaoui2021enhancing}, specifically AA-EEGNet and AA-CascadeNet. These comparisons are conducted against the DS-GTF model, employing three distinct adjacency matrix initialization techniques. All experiments are conducted using the same configuration and data splits as detailed in \cite{abdellaoui2021enhancing}.

\begin{table}[h]
    \centering
     \caption{Comparison of accuracy and standard deviation between subjects for current and proposed models.}
    \setlength{\tabcolsep}{2.7pt}
    \begin{tabular}{cccccccc}
   \toprule
   & & & \multicolumn{3}{c}{DS-GTF} \\ \cline{4-6}
    \textbf{Model} & AA-EEGNet & AA-CascadeNet & FC-Adj & Thresh-Adj  & TopK-Adj\\
        \hline
        \textbf{Accuracy} & $0.90 \pm 0.08$ & $0.93 \pm 0.06$ & $0.94 \pm 0.05$ & \underline{$0.97 \pm 0.03$} & $0.95 \pm 0.05$\\
        \bottomrule
    \end{tabular}
   
    \label{tab:acc}
\end{table}

From the obtained results in Table \ref{tab:acc}, one can notice that the proposed DS-GTF model outperforms the baselines models. This is attributed to the combined strength of a graph-based approach in the spatial stream and a transformer encoder in the temporal stream, which effectively learn rich temporal and spatial representations of the MEG data. Furthermore, employing the TopK-Adj ($K=3$) method for initializing the adjacency matrix leads to a noticeable boost in test accuracy.

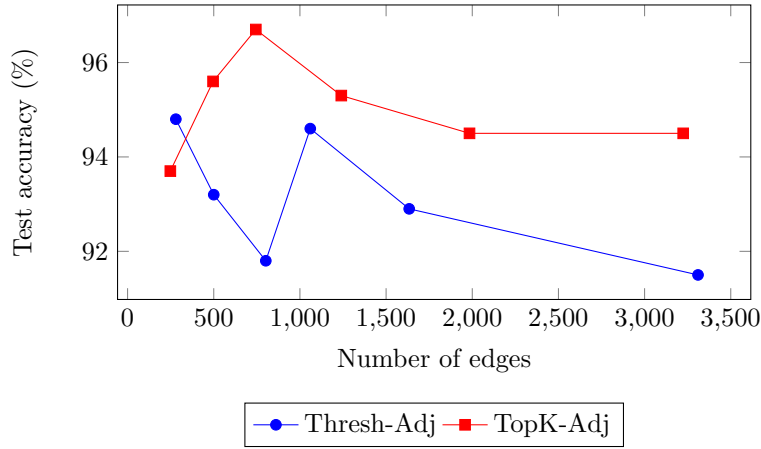
\begin{figure}[h]
  \centering
    \begin{tikzpicture}
      \begin{axis}[
        xlabel={Number of edges},
        ylabel={Test accuracy (\%)},
        width=10cm,
        height = 5.5cm,
        legend style={at={(0.5,-0.35)}, anchor=north,legend columns=-1},
        ]
        
        \addplot[mark=*,blue] coordinates {
          (280,94.8)
          (500,93.2)
          (802,91.8)
          (1060,94.6)
          (1634,92.9)
          (3310,91.5)
        };
        \addlegendentry{Thresh-Adj}
        
        \addplot[mark=square*,red] coordinates {
        (248, 93.7)
        ( 496, 95.6)
        ( 744, 96.7)
        ( 1240, 95.3)
        ( 1984, 94.5)
        ( 3224, 94.5)
        };
        \addlegendentry{TopK-Adj}        
      \end{axis}
    \end{tikzpicture}
    \caption{Test accuracy comparison between TopK-Adj and Thresh-Adj.}
    \label{fig:comparison}
\end{figure}

The comparison displayed in Fig. \ref{fig:comparison} reveals the performance characteristics of the TopK as well and Thresholded adjacency matrix initialization methods in relation to different parameter values. For TopK-Adj a larger $K$ results in more edges while for Thresh-Adj a smaller thresholds results in more edges. These methods are compared by number of generated edges, to observe how well both methods construct the adjacency matrix in equal scenarios. TopK-Adj performs best with smaller $K$ values, capturing localized brain activation effectively and avoiding noise from excessive connections. In contrast, Thresh-Adj struggles with fewer than $1000$ edges, mainly because this results in a graph where some nodes are unconnected. When more edges are initialized, and connectivity is ensured, this method does improve but still lacks behind TopK-Adj.

\section{Conclusion and Future work}
In this paper the Dual Stream Graph-Trans Fusion Networks (DS-GTF) is introduced for classifying task based MEG data. The model exploits the strength of a graph-based approach in the spatial stream and a transformer encoder in the temporal stream, to learn rich temporal and spatial representations of the MEG data. Furthermore, various techniques for initializing the adjacency matrix of the constructed graphs are investigated. The proposed model demonstrates superior generalization compared to its predecessors, with a notable enhancement in performance for classifying brain states and a reduction in the respective standard deviation across multiple test subjects. Moreover, this method has achieved good accuracy without the need of domain knowledge, making MEG decoding a more approachable field for data practitioners.

\begin{footnotesize}



\bibliographystyle{abbrv}
\bibliography{cas-refs}





\end{footnotesize}

\end{document}